\documentclass[aps,prl,twocolumn,superscriptaddress,showpacs]{revtex4}

\usepackage{graphicx}
\usepackage[ansinew]{inputenc}

\usepackage{color}  

\newcommand{\gno}[1]{{\sffamily\textbf{#1}}}

\begin{document}
\title{Shifting the Voltage Drop in Electron Transport through a Single Molecule}

\author{Sujoy Karan}
\email{karan@physik.uni-kiel.de}
\affiliation{Institut f\"ur Experimentelle und Angewandte Physik, Christian-Albrechts-Universit\"at zu Kiel, 24098 Kiel, Germany}

\author{David Jacob}
\affiliation{Max-Planck-Institut f\"ur Mikrostrukturphysik, Weinberg 2, 06120 Halle, Germany}

\author{Michael Karolak}
\affiliation{Institut f\"ur Theoretische Physik und Astrophysik, Universit\"at W\"urzburg, Am Hubland, 97074 W\"urzburg, Germany}

\author{Christian Hamann}
\affiliation{Institut f\"ur Experimentelle und Angewandte Physik, Christian-Albrechts-Universit\"at zu Kiel, 24098 Kiel, Germany}

\author{Yongfeng Wang}
\altaffiliation{Present address: Key Laboratory for the Physics and Chemistry of Nanodevices, Department of Electronics, Peking University, Beijing 100871, P.~R.~China}
\affiliation{Institut f\"ur Experimentelle und Angewandte Physik, Christian-Albrechts-Universit\"at zu Kiel, 24098 Kiel, Germany}

\author{Alexander Weismann}
\email{weismann@physik.uni-kiel.de}
\affiliation{Institut f\"ur Experimentelle und Angewandte Physik, Christian-Albrechts-Universit\"at zu Kiel, 24098 Kiel, Germany}

\author{Alexander I. Lichtenstein}
\affiliation{I.~Institut f\"ur Theoretische Physik, Universit\"at Hamburg, 20355 Hamburg, Germany}

\author{Richard Berndt}

\affiliation{Institut f\"ur Experimentelle und Angewandte Physik, Christian-Albrechts-Universit\"at zu Kiel, 24098 Kiel, Germany}

\begin{abstract}

A Mn-porphyrin was contacted on Au(111) in a low-temperature scanning tunneling microscope (STM). 
Differential conductance spectra show a zero-bias resonance that is due to an underscreened Kondo effect according to many-body calculations.
When the Mn center is contacted by the STM tip, the spectrum appears to invert along the voltage axis.
A drastic change in the electrostatic potential of the molecule involving a small geometric relaxation is found to cause this observation.
\end{abstract}


\maketitle

The spin states of molecules at surfaces are intriguing because of the interaction of localized molecular orbitals with delocalized substrate states.
When a molecular spin is screened by delocalized electrons, the Kondo effect may be observed, which gives rise to a characteristic anomaly in the excitation spectrum close to the Fermi energy \cite{molkon1, hla, aitor, maki, komeda, dilullo}.
Several approaches are being explored to control molecular Kondo systems at surfaces.
These include the application of electric fields \cite{meded_electrical_2011}, mechanical deformation \cite{parks_mechanical_2010}, chemical modification of the ligand shell \cite{Wae, CO, FeTPP, Liu}, and electronic excitation followed by spin state trapping \cite{SCO, SCOWU}.  

In purely metallic systems, the proximity of a metal tip in a scanning tunneling microscope (STM) has been shown to modify the Kondo effect through changes of the Co $d$ band and exchange interaction \cite{CoContact, FeCoContact, Bjo}.
It seems natural to extend this method to the spin states of metal complexes using the tip as an additional ligand.
By moving the tip, the ligand strength may be continuously modified with the subsequent changes in the local environment of the molecule, making this approach particularly appealing.
The concomitant forces were observed to lead, in extreme cases, to the transfer of an atom or molecule from the sample to the tip \cite{eigler, schull}.

Moreover, the electrostatic potential at the molecule may be modified.
Although the electrical field in a nanoscopic junction is fundamentally interesting and drives current through the molecule little is known about it from experiments.
Usually, it is assumed that the states of an adsorbed molecule are pinned with respect to the Fermi level of the substrate.
A voltage drop between molecules and their substrate has been reported in cases where the molecules or parts of them were decoupled from a metal substrate by insulating layers~\cite{Nazin, li_single-electron_2006, cyrus_nnano} or by a particular geometric arrangement such as rod-shaped molecules adsorbed vertically \cite{Reif} or molecular double deckers \cite{Matino}.  
The question of the voltage drop within a molecular junction has been addressed by nonequilibrium \textit{ab initio} quantum transport calculations.
These calculations show that most of the voltage drops at the weakest bond of the junction~\cite{Liang_2004,Bevan_2008}.
Calculations also indicate that a voltage drop between molecule and substrate may lead to negative differential resistance \cite{arxi} and two-level fluctuations \cite{Brandbyge_PRB} of C$_{60}$ junctions as well as rectification in  a molecular wire \cite{taylor}.

Here we present STM data and results of {\it ab initio} calculations for a metal complex on Au(111). 
In the limit of weak interaction with the STM tip the complex exhibits an underscreened high-spin Kondo effect.
When the tip is moved closer to the metal ion, the differential conductance spectrum changes drastically.
It turns out, however, that this change is due to sub-Angstrom structural relaxations.
Similar to a mechanically operated switch, the voltage drop over the junction moves from the tip-molecule gap to the molecule-substrate bond.

Purified 5, 10, 15, 20-tetrakis(4-sulfonatophenyl)-21H, 23H-porphine manganese(III) chloride (MnTPS-Cl) was dissolved in aqueous methanol in the presence of 1 vol$\%$ acetic acid. 
Mass-selected  singly charged MnTPS cations originating from the dissolution of Cl ions were electrosprayed onto clean Au(111) surfaces in ultrahigh vacuum (UHV) \cite{chh}. 
Thus prepared samples were cooled to 77~K and subsequently transferred into a STM, which was operated at $\sim5$~K in UHV\@.
Tips were electrochemically etched from tungsten wire and further prepared \textit{in situ} by indentation into the substrate.
A sinusoidal modulation (2~mV$_\mathrm{rms}$, 1.2~kHz) was added to the bias to record spectra of the differential conductance.

\begin{figure}[htp]
  \includegraphics[width=0.9\linewidth]{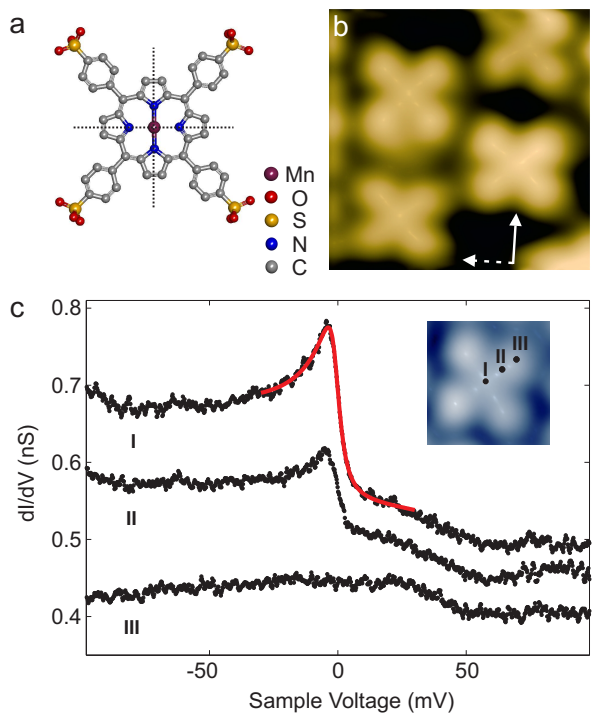}
  \caption{
    (a) Structure of MnTPS (H atoms not shown).
    Dotted lines through pairs of opposite pyrrole units indicate $C_2$ axes.
    (b) STM topograph ($4 \times 4$~nm$^2$, sample voltage $V= 2.65$~V, current $I= 50$~pA) of MnTPS agglomerated into an island on Au(111).
    The data are displayed in a pseudo-three-dimensional fashion.
    The apparent height of the molecular lobes is $\approx 0.28$~nm.
    Solid and dashed vectors indicate the $\left[0\bar{1}1\right]$ and $\left[\bar{2}11\right]$ directions of Au(111), respectively.
    (c) Spatial evolution of $dI/dV$ spectra of MnTPS\@.
    Dots in the inset indicate the positions of the measurements (spectrum \gno{I} above the molecular center, spectrum \gno{III} at the center of a lobe).
    For clarity, spectra \gno{II} and \gno{III} are shifted by $-0.05$ and $-0.1$~nS, respectively. 
    The solid red line depicts a Frota fit to spectrum \gno{I}\@.
  }
  \label{pic}
\end{figure}

Figure~\ref{pic}(a) schematically shows MnTPS, which hosts a Mn ion at the center of a tetraphenylporphyrin with four sulfonyl hydroxide (--SO$_3$H) groups occupying \textit{para} positions of the phenyl rings.
When adsorbed on Au(111), the molecules appear quatrefoil shaped in STM topographs at positive sample voltage $V$ [Fig.~\ref{pic}(b)], as typically observed for porphyrins and phthalocyanines \cite{maki, aitor, FeTPP}.
One of the molecular C$_2$ axes preferably aligns along a $\left<0\bar{1}1\right>$ direction of Au(111) within the agglomerates. 
However, deviations of $\approx \pm 5^\circ$ from this orientation were observed.
Slight variations in apparent heights among the lobes of a particular as well as of different molecules are likely due to incommensurability with the substrate lattice.
This structural variability may indicate a fairly weak molecule-substrate interaction as expected for Au.

\begin{figure}[htp]
  \includegraphics[width=0.82\linewidth]{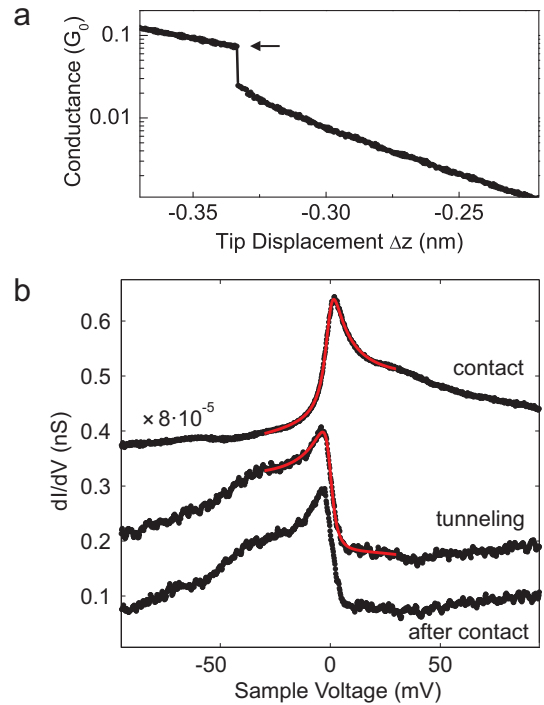}
  \caption{
    (a) Conductance vs displacement $\Delta z$ of the tip from the tunneling condition ($\Delta z=0$ at $V=100$~mV and $I=0.05$~nA) towards contact at the center of a MnTPS molecule.
    At contact formation (arrow) the conductance is 0.074~$G_0$ with $G_0 = 2e^2/h$ being the conductance quantum.
    The conductance measured from different molecules scattered in the range of $0.07 \pm 0.03 G_0$. 
    (b) $dI/dV$ spectra acquired at $\Delta z = 0$ (tunneling) and 0.33~nm (contact) at the center of a molecule.
    The bottom two curves are shifted by $-0.45$ and $-0.35$~nS, respectively, for clarity.
    Solid red lines show fits of Frota functions in the range of $\pm 30$~mV\@.
  }
  \label{iz}
\end{figure}

$dI/dV$ spectra measured over various positions of the molecules are shown in Fig.~\ref{pic}(c).
Above the Mn center (spectrum \gno{I}) a resonance is observed close to zero bias, which continuously vanishes as the tip is moved sideways onto a lobe (spectra \gno{II} and \gno{III}).
Following previous reports from metals \cite{konrev} and metal porphyrins \cite{hla, Xue} we attribute the resonance as a signature of the Kondo effect.
The spectrum of the Mn center may be described well by a Frota line \cite{Frota, pruser_mapping_PRL} (solid red line) leading to an estimated Kondo temperature $T_K \sim100$~K 
\footnote{Different definitions of the Kondo temperature exist in the literature. For the sake of comparability to the Kondo temperatures extracted from previous STM experiments we use the definition that relates $T_K$ to the half width at half maximum of the symmetric resonance. This gives $k_BT_K=2.54\Gamma$, where $k_B$ is the Boltzmann constant and the width $\Gamma$ of the Frota line is defined in REf.~\cite{pruser_mapping_PRL}.}.
$T_K$ measured from different molecules scattered in the range $100(7)$~K\@.
No systematic effect of the position of a molecule in the agglomerates on the line shape was observed.   

Next, the tip was brought closer to the center of a molecule until contact was reached.
Data recorded for a molecule in a fcc region of the Au(111) reconstruction are shown in Fig.~\ref{iz}(a).
$\Delta z$ is the displacement of the tip from its initial height defined by  $V=100$~mV and $I=0.05$~nA\@.
The abrupt rise of the conductance at $\Delta z = 0.33$~nm signals the transition to contact.  

$dI/dV$ spectra were acquired by holding the tip at predetermined heights over the molecular center [Fig.~\ref{iz}(b)].
Spectra recorded in the tunneling range exhibit similar line shapes (only spectra corresponding to $\Delta z=0$ are shown).
At contact, a resonance near zero bias is still observed but surprisingly the line shape appears to be inverted on the voltage axis.
Further approach of the tip from the point of contact does not change the shape of the spectrum.
Approach of the tip beyond $\sim 0.03$~nm from the point of contact usually led to the destruction of the junction.
Very similar Kondo temperatures ($\sim 100$~K) are obtained from the fits of Frota line shapes to both the tunneling and contact spectra. 
Tunneling spectra ($\Delta z=0$) recorded after the contact experiment were unaltered [bottom curve in Fig.~\ref{iz}(b)], excluding the possibility that an irreversible change of the tip or the molecule occurred at contact. 
Isolated molecules, which were prepared by manipulation with the tip, exhibited tunneling characteristics 
virtually identical to that of the molecules in the agglomerates [Fig.~\ref{iz}(b)].
However, at contact isolated molecules were usually unstable.

At first glance, it may seem natural to attribute a changed Kondo line to a modified molecular spin state. 
On the other hand, it appears difficult to interpret the observed inversion of the line in such a scenario. 
We therefore performed density functional theory (DFT) based \textit{ab initio} electronic structure and transport calculations to 
investigate to what extent the Kondo physics is modified by the proximity of the tip.
The calculation explicitly takes into account the strong dynamic correlations originating from the Mn $3d$-shell 
that give rise to the Kondo effect \cite{Jacob:PRL:2009,Karolak:PRL:2011,Jacob:PRB:2013}. 
First, the MnTPS molecule was relaxed in the junction between the Au(111) substrate and the STM tip, modeled by 
a small Au pyramid built in the [111] direction, using \texttt{VASP}~\cite{vasp1, vasp2} with the exchange-correlation 
functional by Perdew, Burke and Ernzerhof~\cite{PBE} 
and the dispersion correction due to Grimme~\cite{DFT-D3} (see details in the Supplemental Material~\cite{supp}).
Figures~\ref{fig:theory}(a) and \ref{fig:theory}(b) show the resulting structures in the tunneling regime and at contact, respectively.
Without the tip, the Mn center is found to be at a height of 2.77~\r{A} above the surface.
When the tip is brought closer to the Mn center, the molecule is lifted from the surface.
At contact (tip at a height of 6~\r{A} above the surface), the displacement of Mn center is 
$\approx 0.7$~\r{A}~\footnote{At tip heights exceeding 6~\r{A} above the substrate the molecule was not lifted further.
We thus define 6~\r{A} as the point of contact.}.

\begin{figure}
  \includegraphics[width=\linewidth]{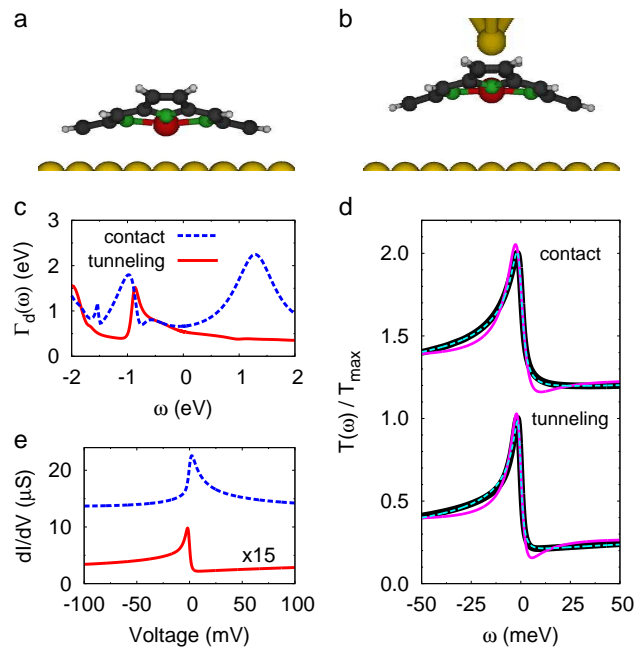}
  \caption{
    \label{fig:theory}
    (a),(b) Schematics of relaxed structures in tunneling and contact regimes. At contact, 
    which corresponds to a tip height of 6~\r{A} above the substrate, the Mn center is lifted 
    from the surface by 0.7~\r{A} with respect to the tunneling regime. 
    This vertical displacement is exaggerated in the figure for better perceptibility.
    (c) Broadening of $3d_{z^2}$ orbital in the tunneling (solid red line) and contact (dashed blue line) regimes.
    (d) Calculated transmission functions $T(\omega)$ normalized to the maximum transmission $T_{\rm max}$ 
    in the tunneling and contact (solid black lines) regimes along with fits using Frota (dashed cyan lines) and Fano (solid magenta lines) line shapes. Fits minimize the mean-square deviation over the interval of $–100$ to 100~meV. For clarity, a narrower range is shown.
    The transmission curve at contact is shifted by 1.
    (e) Calculated $dI/dV$ curves in the tunneling (solid red line) and in the contact regime (dashed blue line). 
    $\alpha$ is 0.02 and 0.8 in the tunneling and contact regimes, respectively. The  contact spectrum is shifted by 10~$\mu$S.
  }
\end{figure}
 
In order to determine the dynamic single-particle broadening $\Gamma_{d}(\omega)$ of the Mn $3d$-levels due to the coupling to the organic framework of the molecule, the substrate, and the tip, DFT-based transport calculations were performed using the \texttt{ANT.G} code~\cite{Jacob:JCP:2011}.
To make the calculations computationally feasible we replaced the sulfonatophenyl groups of the molecule by hydrogen atoms, which leaves the local environment of the Mn center unchanged.
Figure~\ref{fig:theory}(c) shows $\Gamma_{d}(\omega)$ of the $3d_{z^2}$ orbital, which is the only orbital that couples directly to the $s$-type conduction electrons of the substrate (similar to MnPc \cite{Jacob:PRB:2013, Kuegel:NL:2014}) and effectively binds to the $s$ level of the Au tip atom (details in the Supplemental Material~\cite{supp}).
Hence, the $3d_{z^2}$-orbital is essentially the only $d$-orbital that responds to the lifting of the molecule from the substrate.
In the tunneling regime the broadening is featureless around the Fermi energy $E_F$ [Fig.~\ref{fig:theory}(c), solid red line], while the coupling to the $s$-level of the tip at contact leads to a broad feature at $\approx 1.2$~eV above $E_F$ [Fig.~\ref{fig:theory}(c), dashed blue line].
The concomitant increased broadening at $E_F$ is partially compensated by reduced coupling to the substrate. 

To capture correlation effects beyond the mean-field approximation, the Mn $3d$ shell was augmented by a Hubbard interaction term that takes into account density-density interactions and Hund's rule coupling.
We assume an intraorbital repulsion $U=4.5$~eV, an interorbital repulsion $U^\prime=3.1$~eV, and a Hund's coupling $J_H=0.7$~eV
\footnote{These values are slightly smaller than the ones used for MnPc \cite{Jacob:PRB:2013, Kuegel:NL:2014}.
However, the results are hardly affected by changes of these parameters by up to 30\%.}. 
The one-crossing approximation (OCA)~\cite{Haule:PRB:2001} was used to solve the multiorbital Anderson impurity model of the Mn $3d$ shell coupled to the rest of the system.
The solution yields the spectral function of the Mn $3d$-shell $A_d(\omega)$ (see Supplemental Material~\cite{supp}).
Only the $3d_{z^2}$ orbital couples to the conduction electrons near $E_F$ and thus shows a Kondo resonance.
While the width of the resonance is comparable to that of the $dI/dV$ spectra recorded in the tunneling regime, a slightly broader resonance appears at contact due to a small increase of $\Gamma_{d}(E_F)$ and charge fluctuations of the $3d_{z^2}$ orbital (see Supplemental Material~\cite{supp}).
The total spin of the Mn $3d$ shell is found to be approximately 1.7, i.e., halfway between spin-3/2 and spin-2 systems. 
Hence MnTPS displays an underscreened high-spin Kondo effect.

To obtain the low-bias transport properties of the system we calculated the transmission function $T(\omega) = {\rm Tr}[\hat\Gamma_{\rm T}(\omega)\hat{G}_{\rm M}^\dagger(\omega)\hat{\Gamma}_{\rm S}(\omega)\hat{G}_{\rm M}(\omega)]$, where $\hat{G}_{\rm M}(\omega)$ is the correlated Green's function of the extended molecule that contains the Kondo resonance.
$\hat\Gamma_{\rm T}$ and $\hat{\Gamma}_{\rm S}$ are operators describing the coupling to the tip and the substrate, respectively.
Figure~\ref{fig:theory}(d) shows the calculated transmission curves (for both the tunneling and contact regimes) that exhibit asymmetric resonances similar in shape to those of the $dI/dV$ spectra measured in the tunneling regime (Fig.~\ref{iz}(b)). 
However, in contrast to the experiment no line shape inversion occurs at contact.
Figure~\ref{fig:theory}(d) also shows fits of Fano (solid magenta lines) and Frota (dotted cyan) functions to the calculated transmissions for both tunneling and contact. 
Interestingly, the fit of the Frota line to our results, which are based on the OCA, is superior. 
Similar observations were previously made for spectral functions calculated using numerical renormalization group~\cite{nu_renorm_group, pruser_long-range_2012} and quantum Monte Carlo~\cite{quan_monte_carlo} approaches.
Second order perturbation theory, on the other hand, results in a Fano line shape~\cite{lorentzian}.

The difference in line shape between the calculated transmission functions and the measured $dI/dV$ is understood by considering the voltage drop across the junction. 
The current through the molecule can be written in terms of the transmission function as $I(V) = \frac{2e}{h}\int_{\mu}^{\mu+eV} d\omega \, T(\omega-\alpha eV)$, where $\mu$ is the chemical potential of the substrate and $\alpha$ is a dimensionless parameter determining the local chemical potential at the position of the molecule \cite{Reif, Wu_PRL_2004}.
Similar to a voltage divider, $\alpha$ can be obtained from the broadenings $\Gamma_t$ ($\Gamma_s$) due to the tip (substrate) as $\alpha=\frac{\Gamma_t}{\Gamma_t+\Gamma_s}$. 
Since the transport through MnTPS is dominated by transmission through the $3d_{z^2}$ orbital, for $\Gamma_t$ and $\Gamma_s$ we use the couplings of the Mn $3d_{z^2}$ orbital to the tip and substrate, respectively.
We estimate $\alpha=0.02$ with $\Gamma_t\sim0.01$~eV and $\Gamma_s\sim0.54$~eV in the tunneling regime, 
and $\alpha=0.7$ at contact (tip 6~\r{A} above the surface) with $\Gamma_t\sim0.47$~eV and $\Gamma_s\sim0.2$~eV.
As expected, e.~g., from nonequilibrium quantum transport calculations~\cite{Bevan_2008}, 
$\Gamma_t\ll\Gamma_s$ in the tunneling range, $\alpha$ is small, the potential at the molecule is 
pinned to $\mu$, and the applied bias drops over the vacuum gap between the tip and the molecule. 

Figure~\ref{fig:theory}(e) shows the calculated $dI/dV$ spectra for $\alpha=0.02$ (tunneling) and $\alpha=0.8$ (contact).
The latter value is chosen because the estimated $\alpha=0.7$ gives a rather symmetric resonance (also see Supplemental Material~\cite{supp}). 
Like in the experiments, the line shape of the spectrum at contact (dashed blue line) is inverted with respect to the tunneling regime (solid red line).
Since no such inversion occurs in the underlying transmission curves [Fig.~\ref{fig:theory}(d)], the observed inversion is solely due to the modified local potential of the molecule.
In other words, the voltage drop has shifted to the molecule-substrate spacing \footnote{
Reference~\cite{Wu_PRL_2004} pointed out this possibility. 
However, no corresponding experimental observation was made due to junction instability at high conductances.}.
This shift is not only due to an increased molecule--tip coupling but also to a sub-Angstrom displacement of the molecule, which reduces $\Gamma_s$.
Moreover, since $\alpha \ge 0.8$ is required to produce an asymmetry comparable to that of the experimental data, we conclude that at least 80\% of the voltage drops between molecule and substrate in the contact regime.

A second, less obvious difference between tunneling and contact is that the Kondo screening of the molecular spin is mainly mediated by the electron systems of the substrate and the tip, respectively.
In the present experiments, the tip was covered with Au, the material of the substrate and the Kondo temperature happens to be hardly changed.
Occasionally, with some tips, the Kondo resonance disappeared in the contact range, possibly because the material at the apex of those tips did not provide the required screening. 

The model of the voltage drop used above has been very useful in interpreting data from double-barrier situations like tunneling to a molecule on a thin insulator.
The data presented here for a flat molecule between two metal electrodes, shows that the voltage drop can be very relevant ($\ge 80$\% between substrate and molecule) in a case that is less expected.
While for MnTPS ($\alpha$ close to one) an inversion of the line shape is observed, a more symmetric coupling ($\alpha \approx 0.5$) would cause an artificial broadening of the measured resonance.
In summary, the electrical potential in molecular and atomic junctions apparently deserves more attention.

Financial support of the Deutsche Forschungsgemeinschaft (DFG) via Sonderforschungsbereich 677 and Forschergruppe 1162 and 1346 is acknowledged.

\appendix

\section{Supplemental Material}

\subsection{Structure relaxations}

\begin{figure*}
  \begin{center}
    \includegraphics[width=0.95\linewidth]{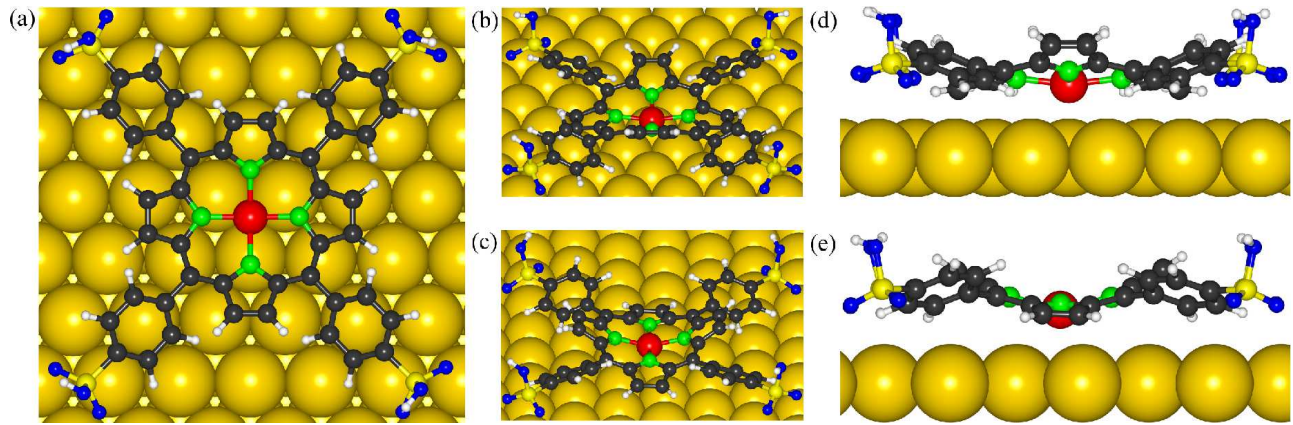}
  \end{center}
  \caption{\label{fig:struct}
    Relaxed structures of MnTPS molecule on Au(111) in the tunneling regime. (a) Top view of the molecule adsorbed on the \textit{fcc} hollow site. (b,c) Side view along the two N-Mn-N axes with a $30^\circ$ inclination w.r.t. the Au surface. (d,e) Side view along the two N-Mn-N axes. For subsequent DFT+many-body calculations the MnP ``core" of the molecule has been used.
  }
\end{figure*}

The structures of an MnTPS molecule have been relaxed on a Au(111) substrate using VASP [38, 39], 
both in the absence and in the presence of a Au tip (in form of a pyramid grown in the [111] direction) in the proximity of the central Mn atom. 
We have used the Perdew-Burke-Ernzerhof (PBE) functional [40] and included dispersion correction due to Grimme (DFT-D3) [41]. 
The experimental situation has been modeled using a 3 layer slab of Au(111) extending over $10\times{}10$ lateral primitive unit cells. For the relaxations we used the $\Gamma$ point in the Brillouin zone only. All crystal structures have been relaxed until the forces acting on each atom were smaller than 0.05~eV/\r{A}. 
The planar geometry of the molecule is strongly deformed on the surface, with the porphine ring forming a saddle structure. 
The phenyl groups rotate themselves out of the molecular plane by about $30^\circ$ as shown in Fig. \ref{fig:struct}. 
Energetically the \textit{fcc} hollow sites are found as the most favorable positions for the molecular center.
In the absence of a tip, the distance of the Mn center from the Au surface is 2.77~\r{A}. 
When the tip is brought into the proximity of the Mn center, the molecule is lifted from the surface. 
For the apex atom at the heights of 6.0, 5.8, and 5.6~\r{A} above the surface, the Mn center is lifted 
by about 0.7, 0.5, and 0.3~\r{A}, respectively, from its initial position.

\subsection{DFT+OCA electronic structure and transport calculations}

We have performed \textit{ab initio} electronic structure and transport calculations explicitly taking 
into account the strong dynamic correlations originating from the 
Mn $3d$-shell that give rise to the Kondo effect, as described in previous work [35-37].
Using the ANT.G \emph{ab initio} transport code [42] we have first performed DFT calculations of an embedded cluster containing the molecule, a finite part of the Au surface and the pyramidal Au-tip.
Both the tip and the finite Au surface have been embedded into electrodes modeled by tight-binding Bethe lattices with realistic tight-binding parameters for bulk Au obtained from DFT calculations. 
In order to make the calculations computationally feasible for the ANT.G code, we replaced the sulfonatophenyl groups of the molecule by hydrogen atoms, leaving the Mn center in a local environment similar to that of the original structure.

\begin{figure}
  \begin{center}
    \includegraphics[width=0.9\linewidth]{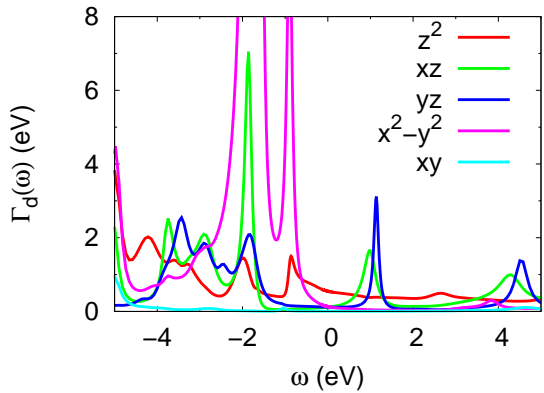}
  \end{center}
  \caption{\label{fig:gamma}
    Orbital resolved dynamic broadening $\Gamma_d(\omega)=-{\rm Im}\Delta_d(\omega)$ for Mn $3d$-shell in the tunneling regime.
  }
\end{figure}

From the converged DFT calculation we obtain the Kohn-Sham Green's function of the embedded cluster,
\begin{equation}
  \hat{G}_{\rm C}^0(\omega) = [\omega+\mu-\hat{H}_{\rm C}^0-\hat\Sigma_{\rm T}(\omega)-\hat\Sigma_{\rm S}(\omega)]^{-1},
\end{equation}
where $\mu$ is the chemical potential, $\hat{H}_{\rm C}^0$ is the Kohn-Sham (KS) Hamiltonian of the 
cluster which yields a mean-field description of the electronic structure of the cluster
and $\hat\Sigma_{\rm T}(\omega)$ and $\hat\Sigma_{\rm S}(\omega)$ are the embedding self-energies
which describe the coupling of the cluster to the bulk electrodes of the STM and substrate,
respectively.
From $\hat{G}_{\rm C}^0(\omega)$ we calculate the so-called hybridization function of the Mn $3d$-shell,
\begin{equation}
  \hat\Delta_{d}(\omega) = \omega+\mu-\hat{h}^0_{\rm d}-[\hat{G}_{d}^0(\omega)]^{-1},
\end{equation}
where $\hat{h}_{d}^0$ and $\hat{G}_{d}^0(\omega)$ are the projected KS Hamiltonian and KS Green's function, respectively, for the Mn $3d$ states. 
The negative imaginary part of $\hat\Delta_{d}(\omega)$ yields the (dynamic) single-particle broadening $\hat\Gamma_{d}(\omega)$ of the $3d$-levels due to the coupling to the rest of the system comprising the organic rest of the molecule, substrate and the tip. 
Figure~\ref{fig:gamma} shows the orbitally resolved single-particle broadening $\hat\Gamma_{d}(\omega)$ for the Mn $3d$-shell in the tunneling regime. 
The hybridization functions resemble 
the ones calculated for manganese phthalocyanine (MnPc) on Pb(111) and Ag(001) [37, 43], but differ in some important details.
The broadening of the $3d_{z^2}$-orbital is relatively featureless around the Fermi level indicating a coupling of that orbital predominantly to the $s$-type conduction electrons in the Au substrate. 
Near the Fermi level the $d_{xz}$- and $d_{yz}$-orbitals do not couple directly to the substrate, but the appearance of Lorentzian peaks in the hybridization functions indicate indirect coupling of those orbitals to the substrate via organic ligands. In contrast to MnPc, these peaks are not directly at the Fermi level but about 1~eV above. 
The $d_{x^2-y^2}$ orbital has an extremely strong coupling to the organic rest of the molecule at $-2$~eV which leads to quenching of that orbital, while the $d_{xy}$-orbital does not couple at all to the substrate or the organic rest for low energies.

In order to capture dynamic correlation effects originating from the strongly interacting $3d$-electrons of the Mn center, the Mn $3d$-shell is augmented by a Hubbard type interaction term,
\begin{equation}
  \mathcal{H}_U = \frac{1}{2} \sum_{ijkl\sigma\sigma^\prime} U_{ijkl} d_{i\sigma}^\dagger d_{j\sigma^\prime}^\dagger d_{l\sigma^\prime} d_{k\sigma},
\end{equation}
where $U_{ijkl}$ are the matrix elements of the effective Coulomb interaction for the Mn $3d$-shell. The effective Coulomb interaction is much smaller than the bare one due to RPA screening by the conduction electrons in the substrate, tip and organic rest of the molecule. Here we take a simplified model interaction taking into account only density-density interactions $U_{ijij}$ and Hund's rule coupling $U_{ijji}$. We assume an intra-orbital repulsion $U=4.5$~eV, an inter-orbital repulsion $U^\prime=3.1$~eV  and a Hund's coupling $J_H=0.7~eV$. 
Since the Coulomb interaction within the $3d$-shell has already been taken into account on a mean-field level by the Kohn-Sham DFT Hamiltonian $\hat{H}_d^0$, we subtract a double-counting correction (DCC) in order to obtain the bare energy levels of the Mn $3d$-shell: $\hat\epsilon_{d} = \hat{H}^0_{d} - \hat{V}_{dc}$. Here we choose the so-called fully-localized limit (FLL) DCC scheme [51].

\begin{figure}
  \begin{center}
    \includegraphics[width=0.9\linewidth]{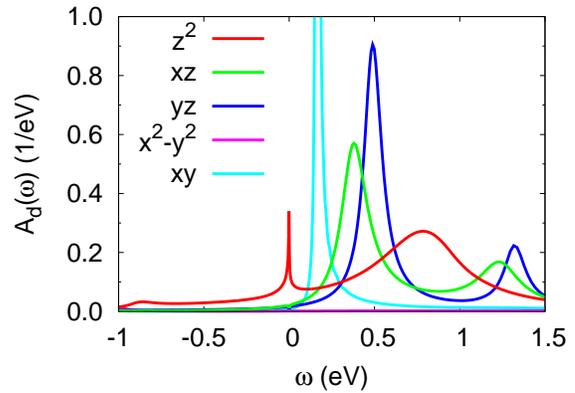}
  \end{center}
  \caption{\label{fig:spectra}
    Orbital resolved spectral function in the tunneling regime.
  }
\end{figure}

The Mn $3d$-levels $\epsilon_{d}$, the effective Coulomb interaction $\mathcal{H}_U$ and the 
hybridization function $\hat\Delta_{d}(\omega)$ completely define a multi-orbital Anderson 
impurity model (AIM). We have used One-Crossing Approximation (OCA) to solve the AIM [44].
The solution of the AIM yields the spectral function of the Mn $3d$-shell $A_d(\omega)$ which is shown in 
Fig.~\ref{fig:spectra} for the tunneling regime. 
Only the $3d_{z^2}$-orbital shows a Kondo peak at the Fermi level.
The other orbitals do not show Kondo peaks but have strong resonances not too far from the Fermi level. 
Overall the Mn $3d$-shell is in a mixed valence situation with a total occupation of $\sim5.5$, although individual orbital-occupancies are rather close to 1.
The total spin of the Mn $3d$-shell is found to be approximately 1.7, \textit{i.e.} half-way between a Spin-3/2 and a Spin-2 systems. The spin in the $d_{x^2-y^2}$-orbital is completely quenched by the strong coupling to the organic ligands.
At contact the $d_{z^2}$-orbital is slightly lowered in energy due to the bonding to the tip atom, increasing the occupancy of that orbital to $\sim1.2$. 

\begin{figure}
  \begin{center}
    \includegraphics[width=0.9\linewidth]{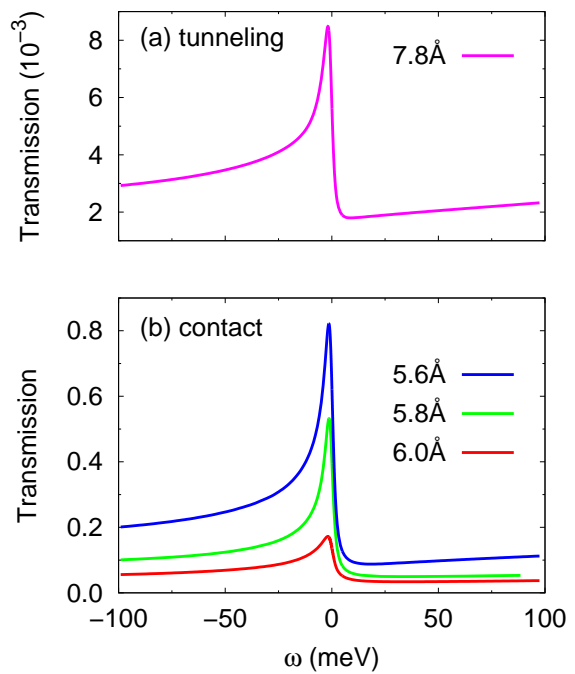}
  \end{center}
  \caption{\label{fig:transm}
    Correlated transmission functions 
    (a) in the tunneling regime with the tip at 7.8\r{A} above the surface, and 
    (b) in the contact regime with tips at different heights above the surface as indicated.
  }
\end{figure}

Using the Kramers-Kronig relation the real part of the correlated $3d$ Green's function $\hat{G}_{d}(\omega)$ has been calculated. From that we obtain the electronic self-energy via $\hat\Sigma_{d}=[\hat{G}_{d}^0]^{-1}-[\hat{G}_{d}]^{-1}$, 
which captures the dynamic correlation effects induced by the effective Coulomb interaction $\mathcal{H}_U$.
The \emph{correlated} electronic structure and transport properties of the whole system is then calculated from the correlated cluster Green's function, 
\begin{equation}
  \hat{G}_{\rm C}(\omega) = [ (\hat{G}_{\rm C}(\omega))^{-1} - \hat\Sigma_{d}(\omega) + \hat{V}_{dc} ]^{-1},
\end{equation}
where the $\hat\Sigma_{d}$ and $\hat{V}_{dc}$ only act on the subspace of the Mn $3d$-shell.
From $\hat{G}_{\rm C}(\omega)$ we calculate the correlated transmission function,
\begin{equation}
  T(\omega) = {\rm Tr}[ \hat\Gamma_{\rm T}(\omega) \hat{G}_{\rm C}^\dagger(\omega) \hat{\Gamma}_{\rm S}(\omega) \hat{G}_{\rm C}(\omega) ],
\end{equation}
which according to Landauer yields the low-bias transport properties, \textit{i.e.} the low-bias differential conductance and current [52]. 
$\hat\Gamma_{\rm T}$ and $\hat{\Gamma}_{\rm S}$ are the imaginary parts of the electrode self-energies
$\hat\Sigma_{\rm T}$ and $\hat{\Sigma}_{\rm S}$, respectively.
Figure~\ref{fig:transm} shows the calculated transmission functions both in the tunneling and in the contact regime for different heights of the tip above the surface. 
For distances below 5.8~\r{A} the transmission function is strongly broadened, \textit{i.e.} the Kondo temperature increases strongly. Also the conductance increases strongly. 
However, junction instability at higher current prohibited to make this observation experimentally.

\begin{figure}[ht!]
  \begin{center}
    \includegraphics[width=0.9\linewidth]{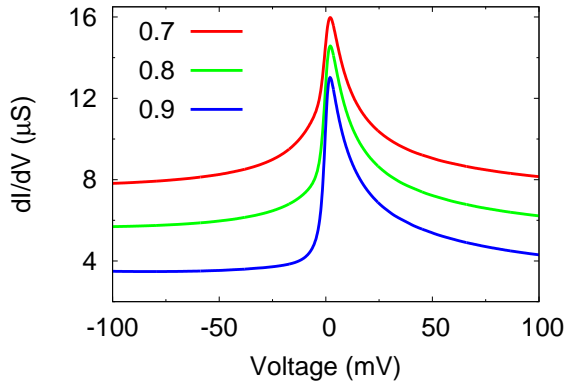}
  \end{center}
  \caption{\label{fig:didv}
    Calculated $dI/dV$ spectra in the contact regime with the tip at 6\r{A} above the substrate
    for different values of the parameter $\alpha$.
  }
\end{figure}

Figure~\ref{fig:didv} shows the calculated $dI/dV$ spectra in the contact regime for different values of $\alpha$, the dimensionless parameter determining the local 
chemical potential at the position of the molecule. 
For $\Gamma_t\ll\Gamma_s$ ($\Gamma_t\gg\Gamma_s$), $\alpha\rightarrow0$ ($\alpha\rightarrow1$) implying that the local potential at the molecule is pinned to the chemical potential of the substrate (the tip).

\end{document}